\documentclass[10pt]{iopart}
\usepackage{graphicx,bbm}
\usepackage{hyperref}
\def\tit#1{}
\def\etal#1{#1}

\def\tr{\,{\rm tr}\,}
\def\ket#1{|#1\rangle}
\def\bra#1{\langle#1|}
\def\ave#1{\langle #1 \rangle}
\def\aveU#1{\langle #1 \rangle_{\rm U}}
\def\aveE#1{\langle #1 \rangle_{\rm E}}
\def\ii{{\rm i}}

\begin{document}

\title{Subsystem dynamics under random Hamiltonian evolution}

\author{Vinayak$^1$ and Marko \v Znidari\v c$^{1,2}$}
\address{$^1$Instituto de Ciencias F\' isicas, Universidad Nacional Aut\' onoma de M\' exico, Cuernavaca, M\' exico}
\address{$^2$Physics Department, Faculty of Mathematics and Physics, University of Ljubljana, Ljubljana, Slovenia}
\date{\today}

\begin{abstract}
We study time evolution of a subsystem's density matrix under unitary evolution, generated by a sufficiently complex, say quantum chaotic, Hamiltonian, modeled by a random matrix. We exactly calculate all coherences, purity and fluctuations. We show numerically that the reduced density matrix can be described in terms of a noncentral correlated Wishart ensemble for which we are able to perform analytical calculations of the eigenvalue density. Our description accounts for a transition from an arbitrary initial state towards a random state at large times, enabling us to determine the convergence time after which random states are reached. We identify and describe a number of other interesting features, like a series of collisions between the largest eigenvalue and the bulk, accompanied by a phase transition in its distribution function.
\end{abstract}

\maketitle
\section{Introduction} 
Isolated systems are usually idealizations that is nevertheless often valid to a good approximation due to small coupling to external degrees of freedom. Frequently though, one would like to describe the dynamics of a subsystem, even if the coupling to ``external'' degrees of freedom is not small. This is of obvious importance for foundations of statistical mechanics where in certain simple situations, for instance in equilibrium, one can derive the stationary distribution of a subsystem under rather general assumptions. Much more difficult, but also more interesting, is the problem of describing the subsystem in a nonequilibrium setting, for instance, its evolution with time. In the present work we achieve just that. 

We completely describe how the statistical properties of a reduced density matrix evolve in time, provided we have a unitary evolution on the whole system. Because we want to avoid system-specific features of a particular Hamiltonian we use a random matrix model. The Hamiltonian of the whole system is given as a random matrix, in our case from a unitary ensemble, although straightforward generalizations are possible also to other symmetry classes. It is well established~\cite{RMT} that random matrix Hamiltonian's are useful to describe certain statistical properties of quantum chaotic systems, meaning that our results should apply to a large class of dynamical systems. For such random matrix model of a reduced density matrix we derive exact analytic results for all coherences as well as for purity. Calculating spectral properties of the reduced density matrix exactly though is still too difficult. With the goal of nevertheless obtaining analytical result in the limit of large systems we statistically describe the reduced density matrix by a Noncentral Correlated Wishart Ensemble (n-CWE)~\cite{Muirhead:82,Letac:04}, i.e., a Wishart Ensemble~\cite{wishart:28} (WE) with nonzero average matrix elements and correlations between rows or columns. Doing this we are able to calculate the eigenvalue density of the reduced density matrix, thereby fully describing all invariant properties of the reduced density matrix at any time. As an example, calculating the average value of the largest eigenvalue we observe a series of its ``collisions'' with the bulk of the spectrum, each time accompanied by a phase transition in its distribution function.

Various Wishart ensembles have been studied in different contexts. The uncorrelated Wishart ensemble with a centered distribution, and its fixed-trace generalization, has received a lot of attention from physicists. Even a partial list of its applications includes diverse fields, like mesoscopic systems~\cite{Beenakker:97}, analysis of time series in quantitative finance~\cite{Laloux:99,Plerou:99} or of EEG signals~\cite{Seba:03}, high-energy physics~\cite{Janik:03} or communication engineering~\cite{Telatar:99}. It describes also properties of random quantum states, important in quantum information theory. A number of interesting phenomena has been discovered, like phase transitions in the distribution of extreme eigenvalues or of Reny entropies~\cite{arul:08,vergassola:09,nadal:10}. Various generalizations of Wishart ensemble have also been studied~\cite{gernot}. On the other hand, n-CWE, being the ensemble that applies to our work, though it appeared early in the literature~\cite{James:64}, is rather unexplored compared to WE. It has been mostly of interest in mathematical statistics~\cite{Muirhead:82,Baik:05,Peche:06,bff} or in signal processing~\cite{speckle}. This work is the first application of n-CWE to physics, thereby opening many new research directions. It turns out that for large matrices correlations are small and one deals with an uncorrelated Noncentral Wishart Ensemble (n-WE) or, due to universality, in some cases with a Correlated Wishart Ensemble (CWE) which has also been of interest in recent years \cite{Simons:04,vinayak,Guhr:10}. Because the natural ensemble in our setting is the n-CWE this should also give a fresh impetus for more detailed mathematical studies of n-CWE. 

Our results also have implications for the much studied quantum protocols to generate random quantum states. An important question is how long does it take to generate a random quantum state. For different random circuit protocols~\cite{Emerson:03}, that can be described as a Markovian process, it has been proved~\cite{circLin} that the convergence time (number of two-qubit operations; one operation per time step) scales with the number of qubits $n$ as $\sim n^2$ (as opposed to previous lower bound~\cite{plenio}). However, from an experimental point of view it might often be easier to implement a chaotic Hamiltonian than a specific fixed random circuit protocol. A proof of convergence time in such setting is still missing, although it has been argued~\cite{znidaric:07} that it should have the same scaling as for random circuits. Our general result provides this missing link, deriving the convergence time for a large class of protocols based on sufficiently complex Hamiltonians. We also show that during evolution the statistical properties of states are well described by the n-WE. One interesting observation is that, because the convergence to the asymptotic invariant distribution is not monotonic, evolving for longer time can, somehow counter-intuitively, worsen the randomness of states. How the spectrum of the reduced density matrix changes with time has been numerically observed in a random protocol~\cite{znidaric:07} and in a dynamical system~\cite{Buch}.

The outline is the following: in Section~\ref{sec:model} we shall present our model for a reduced density matrix and obtain some exact results. In Section~\ref{sec:wishart} we approximate the model from the previous section by a Wishart ensemble, obtaining exact results for the eigenvalue density in the limit of large system size. In Section~\ref{sec:results} we verify theoretical results by numerical simulations. In Sec.~\ref{sec:random} we discuss the implication of our results for the convergence rate towards random states and in Sec.~\ref{sec:conc} we conclude.

\section{The model and averages over the unitary Haar measure}
\label{sec:model}
Let us first define our random matrix model for a reduced density matrix. Dynamics on the whole system is given by a unitary propagator $U^t=\exp{(-\ii H t)}$ acting on a $N\otimes M$ dimensional Hilbert space (without loss of generality we assume $N \le M$). We are interested in the reduced dynamics on an $N$ dimensional factor space obtained by tracing over the ``environment'' of dimension $M$. The state at later time $\ket{\psi(t)}$ is given through the unitary evolution $\ket{\psi(t)}=U^t \ket{\psi(0)}$, and the reduced density matrix is $\rho(t)=\tr_{\rm E}{\ket{\psi(t)}\bra{\psi(t)}}$. The only assumption we need in our exact derivations in this section is that the eigenvectors of $H$ are well described by a random unitary matrix so that we will be able to perform averages over unitary Haar measure of these eigenvectors, denoted by $\aveU{}$, while the eigenvalues of $H$ can be arbitrary. We shall shortly call such a model a random density matrix model (RDMM).

As a concrete example of RDMM, by which we will numerically check our theory, we will use a Hamiltonian from a Gaussian Unitary Ensemble (GUE) in which matrix elements are independent complex Gaussian random numbers~\cite{RMT}. We fix normalization, i.e., the variance of random numbers, by $\ave{|H_{jk}|^2}=\frac{1}{NM}$, giving the asymptotic (large $NM$) spectral span of 4, while the Heisenberg time is $2NM$. Note that for a specific dynamical system normalization will be of course different, however, this results in a trivial rescaling of two relevant time scales: the shortest time scale is the inverse spectral span (the largest energy scale in the system), while the longest, i.e., the Heisenberg time, is given by the inverse of the mean level spacing (the smallest energy scale in the system).  The initial  state on the total system is an arbitrary superposition $\ket{\psi(0)}=\sum_{j,\nu} A_{j,\nu}(0)\ket{j}\ket{\nu}$, where $\ket{j}$ are basis states of a central system and $\ket{\nu}$ are basis states of an ``environment''. The initial density matrix is $\rho^{(0)}=A(0) A^\dagger(0)$. 

We shall now calculate some exact results for the RDMM. Writing $U^t_{j,k}=V_{j,m} {\rm diag}\{\exp{(-\ii E_m t)}\} V^*_{k,m}$ in terms of eigenvectors $V_{j,k}$ and eigenenergies $E_m$, and performing average over unitary Haar-measure distributed eigenvectors $V$ of $H$, we get for the average expansion coefficients $A(t)$ at time $t$,
\begin{equation}
\aveU{A(t)}=f_t\, A(0), \qquad f_t\equiv \frac{1}{NM}\sum_j \exp{(-\ii E_j t)},
\label{eq:ft}
\end{equation}
where $f_t$ depends only on the eigenenergies $E_j$ of $H$. Details of the calculation can be found in the appendix. We could also perform averaging over the spectrum, $\aveE{}$, obtaining the average $\aveE{f_t}\equiv g_t$, and $\aveE{|f_t|^2}\equiv h_t^2$. The eigenvector-average of the density matrix at time $t$ involves averages over monomials consisting of two matrix elements $V_{j,m}$ and two conjugate $V^*_{j,m}$, resulting in (see the appendix)
\begin{equation}
\aveU{\rho(t)}=\frac{(NM)^2 |f_t|^2-1}{(NM)^2-1}\, \rho^{(0)} + \frac{NM^2(1-|f_t|^2)}{(NM)^2-1}\,\mathbbm{1}.
\label{eq:avgrho}
\end{equation}
All averages over the Haar measure of $V_{j,m}$ can be evaluated exactly by the method developed in Ref.~\cite{Collins}. Even though the expression (\ref{eq:avgrho}) is only a starting point for our calculations it has already important consequences. For instance, it gives all coherences (off-diagonal elements of $\rho$) which are of interest in studies of decoherence. From Eq.~(\ref{eq:avgrho}) it is also simple to derive a quantum map $\rho^{(0)} \to \rho(t)$ for our RDMM, which is diagonal in a local basis. This generalizes a recent result~\cite{znidaric:11} for qubits ($N=2$). Furthermore, Eq.(\ref{eq:avgrho}) will be used to define an appropriate Wishart ensemble in Section~\ref{sec:wishart}.

Average purity $I(t)\equiv \aveU{\tr_{\rm E}{[\rho^2(t)]}}$ can also be evaluated exactly in our setting in which the eigenvectors of $H$ are given by a random unitary matrix. The calculation involves averages over products of $4$ elements $V_{j,m}$ and $4$ conjugate ones, see the appendix. The Weingarten function on permutations of $4$ elements needed for such a calculation~\cite{Collins} can be found in Ref.~\cite{znidaric:11}. After straightforward, but tedious calculation, we arrive at the final exact result for the average purity for the RDMM and a product initial state (any $N,M>1$ and $t$),
\begin{equation}
I(t)=I_{\rm r}+B \left( {\cal N}^2 |f_t|^4+{\cal N} v_t+|f_{2t}|^2-4|f_t|^2 \right),
\label{eq:I}
\end{equation}
with $I_{\rm r}\equiv \frac{N+M}{1+NM}$ being the average purity of complex random states~\cite{lubkin:78}, ${\cal N}\equiv NM$, $v_t\equiv [f^2_t]^* f_{2t}+f^2_t [f_{2t}]^*$ and $B\equiv \frac{(N-1)(M-1)}{({\cal N}+3)({\cal N}+1)({\cal N}-1)}$. Purity for the GOE case at certain times has been studied in Ref.~\cite{gorin:02}. Interestingly, for $t\to \infty$ purity does not go towards $I_{\rm r}$ but instead~\cite{footnote1} to $I_{\rm r}+\frac{2(N-1)(M-1)}{{\cal N}({\cal N}+3)({\cal N}+1)}$. For any finite size and any time, even an infinite one, one does not reach perfectly random states! For discrete-time evolution described by circular ensembles this has been observed in Ref.~\cite{Vallejos}.

In a special case when $H$ is from a GUE, and for large ${\cal N}$, we can simplify $I(t)$ even more. The infinite size limit of $f_t$ averaged over GUE eigenenergy spectrum is $g_t= J_1(2t)/t$. In terms of $g_t$ we can write in the leading order in ${\cal N}$ that $h_t^2=g_t^2+{\cal O}(1/{\cal N})$, $\aveE{|f_t|^4}=g_t^4+{\cal O}(1/{\cal N}^2)$, and $\aveE{v_t}=2 g_t^2 g_{2t}+{\cal O}(1/{\cal N}^2)$. Using these we obtain the first three leading orders of purity for GUE Hamiltonian
\begin{equation}
I(t) = g_t^4+\frac{(1-g_t^4)(N+M)}{NM}+\frac{2(g_t^2 g_{2t}-g_t^4)}{NM} + {\cal O}(1/{\cal N}^2).
\label{eq:Iasymp}
\end{equation}

\begin{figure}[ht!]
\includegraphics[angle=-90,width=0.9\textwidth]{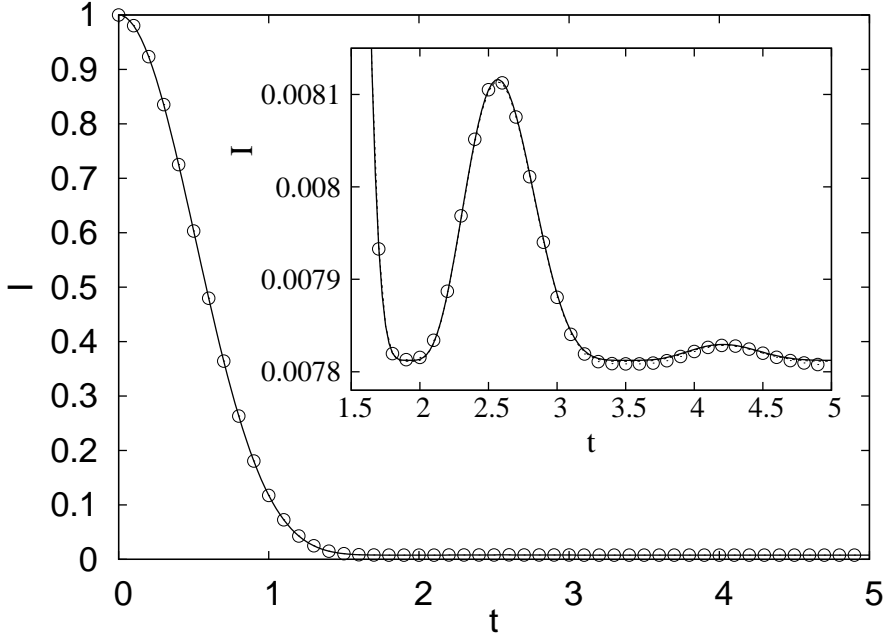}
\caption{Purity for a RDMM, $H \in {\rm GUE}$ and a product initial state. Full curve is the theory, Eq.~(\ref{eq:Iasymp}), circles are numerics; $N=M=256$. Note ``revivals'' of purity around $t \approx 2.5$ and $t\approx 4.2$.}
\label{fig:purk1}
\end{figure}

We shall now check our exact result for purity (\ref{eq:I}) against result of numerical simulation of RDMM for a GUE Hamiltonian. Average purity is shown in Fig.~\ref{fig:purk1}, where one can observe a perfect agreement between the exact purity (\ref{eq:I}) and results of numerical simulation.

We have derived an exact expression for $\aveU{\rho(t)}$ and for the purity in our RDMM. The exact calculation of the spectral density on the other hand still seems to be too difficult. With that in mind we will make one further simplification. To nevertheless be able to obtain analytical results and to give a complete statistical description for the subsystem dynamics we shall in the next section describe the RDMM by a simpler statistical ensemble, namely by a noncentral correlated Wishart ensemble (n-CWE). Such description of the RDMM is only approximate, however, as we shall demonstrate, for large dimensions $N,M$, on which we shall focus in the present work, it becomes exact. The ensemble properties are self-averaging in this limit, meaning that ensemble averaged quantities describe also properties of one particular instance of $\rho(t)$. We leave small size-cases for future studies. 

\section{Wishart ensembles}
\label{sec:wishart}
The idea is to describe the RDMM in an approximate way by a suitable Wishart ensemble. Wishart ensemble parameters are chosen so that the low-order moments are equal to those of the RDMM, for instance, equal to the average reduced density matrix (\ref{eq:avgrho}). Higher order moments of the RDMM and those of the Wishart ensemble will in general differ, however, we expect that the contributions of higher moments become negligible for large system sizes. We shall verify this assumption by numerical simulations. 

\subsection{Generalized Wishart ensembles as a model for RDMM}
The most general Wishart ensemble is n-CWE. Member matrices $W$ of n-CWE are defined as 
\begin{equation}
W\equiv Z Z^\dagger,\qquad Z\equiv Y+\sqrt{\xi} X,
\label{eq:defW}
\end{equation}
through an auxiliary matrix $Z$, in which $Y$ is a fixed matrix determining an offset (``nonzero average of matrix elements'' in n-CWE), and $X$ is an $N\times M$ matrix of independent complex Gaussian variables with zero average and variance $\ave{|X_{jk}|^2}=\frac{1}{NM}$, where $\ave{}$ denotes an ensemble average over Gaussian numbers in $X$. $\xi$ is a positive definite symmetric matrix which takes account for the correlations among the rows of $Z$. Note that we have an ensemble average $\ave{W}=Y Y^\dagger+\frac{1}{N}\xi$. We remark here that initially CWE has been introduced as a model for the covariance matrices where one considers $Y=0$. Additionally, statistical independence of the matrix elements is imposed for the simplicity from technical point of view. On the other hand, for RDMM both conditions are {\it a priory} not necessary, or even not valid (i.e., average matrix elements of $Y$ are nonzero and there are nonzero correlations). These generalizations therefore make n-CWE more suitable candidate for approximating RDMM. 

We now fix parameters of n-CWE by demanding that $Z$ (\ref{eq:defW}) describes $A(t)$ (\ref{eq:ft}) while $W$ should describe density matrix $\rho(t)$ (\ref{eq:avgrho}). Specifically, we have that $\ave{Z}$ equals the average expansion coefficients $\ave{A(t)}$ (\ref{eq:ft}), while the covariance matrix $\xi$ is on the other hand fixed by equating $\ave{W}=\sigma^2 \xi+|g_t|^2 \rho^{(0)}=\aveU{\rho(t)}$, where $\sigma^2=1/N$. This therefore gives us n-CWE parameters
\begin{equation}
Y=g_t\,A(0),\qquad \sigma^2 \xi=\aveU{\rho(t)}-|g_t|^2 \rho^{(0)}.
\label{eq:nCWE}
\end{equation}  
Using Eq.(\ref{eq:avgrho}) we can see that for large $N$ and $M$ correlations are small, that is $\xi$ is proportional to the identity matrix, $\xi = (1-h_t^2)\mathbbm{1}+{\cal O}(1/(NM)^2)$. In the rest of the paper we shall focus on the case of large $N$ and $M$ and we will in turn approximate n-CWE by an uncorrelated n-WE for which $\xi \sim \mathbbm{1}$. 

Even though n-WE is a simpler ensemble than n-CWE compact analytical expressions are still difficult to obtain. For instance, the ensemble averaged eigenvalue density, $\ave{p(x)}$, is not known in a simple form \cite{skap:11}. Fortunately, in some of the cases we discuss below, we can approximate n-WE with that of a simpler CWE. An approximation of using CWE to describe n-WE has been used before~\cite{Kollo:95}. CWE description of the RDMM is obtained by setting
\begin{equation}
Y=0,\qquad \sigma^2\xi=\aveU{\rho(t)}.
\end{equation}
Note that in this case of CWE description correlations $\xi$ are not small because they must also account for a nonzero $\rho^{(0)}$ term in $\rho(t)$ (\ref{eq:avgrho}) which is in n-CWE approximation taken care of by a nonzero $Y$. For most of the parameters and quantities studied in this work we conjecture that the level density is the same for CWE and n-WE, in the leading order. In the absence of rigorous analytical treatment, we leave this conjecture based on some numerical experiments. A non-rigorous argument for the universality, at least for not too small times, is that for the spectral properties of $\rho(t)$ only its moments matter and not for instance expansion coefficients of $\ket{\psi(t)}$, which is fixed by a nonzero $Y$ in n-CWE. Similar universality has also been discovered in some other random matrix ensembles \cite{bff}.  

\subsection{Results from the CWE}

We proceed by repeating analytical results for the eigenvalue density of CWE in the limit of large $N, M$, for more details the reader can consult Ref.~\cite{vinayak}. For a given spectrum of $\xi$ and for large $N,M$, the ensemble averaged density of eigenvalues for CWE, $\ave{p(x)}$, can be determined in terms of its resolvent $G(z)$,
\begin{equation}
\label{res}
\langle{G}(z)\rangle
=
\frac{1}{N}\, \mathsf{Tr}\,
\left(z-\frac{\sigma^{2}}{\kappa}\left[\kappa-1+z\ave{G(z)}\right]
 \xi\right)^{-1},
\end{equation}
where $\kappa=M/N$, as $\ave {p(\lambda)}=\Im{G(\lambda-\ii\,\epsilon)}/\pi$, for positive infinitesimal $\epsilon$. For a simple spectrum of $\xi$ the Eq.~(\ref{res}) can be solved analytically. For instance, for an initial product state one has (again, for large sizes) $\xi_{jk}=\delta_{jk}+(1-\delta_{jk})r$, where $r=h_t^2$. In this case, for large $r$ the largest eigenvalue centered at $\ave{\lambda_1}$ and with Gaussian fluctuations~\cite{Peche:06} around, is separated from the bulk $\lambda_{2,\ldots,N}$~\cite{Baik:05} ($\lambda_j$ are in decreasing order), and equal to
\begin{equation}
\langle\lambda_{1}\rangle=\sigma^{2}\frac{(Nr+1-r)(Nr\kappa+1-r)}{Nr\kappa}.
\label{eq:lam1}
\end{equation}
This equation is valid only if the eigenvalue is separated from the bulk, i.e., if $r > (N\sqrt{\kappa})^{-1}$. It can also be generalized to a case when more than one eigenvalue is separated~\cite{vinayak,bff}. The distribution of eigenvalues of $\rho(t)$ in the bulk is given by the Mar\v{c}enko-Pastur law~\cite{marcenko} with a rescaled variance $(1-r)\sigma^{2}$: 
\begin{equation}
\label{mp}
\langle\,p_{\small {\rm MP}}(\lambda)\rangle=
\kappa\frac{\sqrt{(\lambda_{+}-\lambda)(\lambda-\lambda_{-})}}{2\pi \lambda(1-r)\sigma^{2}},
\end{equation}
where $\lambda_{\pm}\equiv (1-r)\sigma^{2}(\kappa^{-1/2}\pm1)^{2}$, are the minimum and maximum cut-off of the eigenvalue density for the bulk.

As long as the largest eigenvalue is separated from the bulk we can also analytically calculate its variance, though not using the CWE approximation~\cite{footnote2}. If we start with an initial state in which one eigenvalue of $\rho^{(0)}$ is separated from the rest -- the simplest example is an initial product state -- the eigenvector $\ket{x_t}$ corresponding to this eigenvalue is almost the same as $\ket{x_{t=0}}$. Therefore, the variance can be approximately calculated from the variance of the matrix element $\aveU{x_0|\rho(t)|x_0}$. For an initial product state and RDMM such an average can be evaluated exactly (for $N \to \infty$) by the same method as used for the exact calculation of the average purity (\ref{eq:I}). The final result is
\begin{equation}
\sigma^2(\lambda_1) \approx \sigma^2(\rho_{00}) = \frac{2g_t^2+2 g_t^2 g_{2t}-4 g_t^4}{NM} + {\cal O}(\frac{1}{(NM)^2}).
\label{eq:var1}
\end{equation}
Note that for product initial states and short times the largest eigenvalue is much larger than all the others and will therefore govern properties of the reduced density matrix. As we shall see in the next section (e.g., Fig.~\ref{fig:lam01k1}), the largest eigenvalue is separated from the bulk also at times between collisions of the largest eigenvalue with the bulk, that is not necessarily at short times, provided $N$ is large enough (one therefore has to let $N \to \infty$ before letting time to infinity).

\section{Numerical comparison}
\label{sec:results}
In this section we shall verify our theoretical results for the eigenvalue density of $\rho(t)$ for the RDMM, obtained from the approximation by the CWE, Eqs.~(\ref{mp},\ref{eq:lam1}), against results of full numerical simulation of RDMM with GUE Hamiltonian. To obtain RDMM results we expanded short-time propagator $\exp{(-\ii H \Delta t)}$ into a power series, keeping a sufficient number of terms, and acting with it on pure state. Repeating this procedure pure state at a later time is obtained, from which we then calculate the reduced density matrix. For largest $N$'s used the above power-expansion method is faster than directly diagonalizing $H$. We shall also compare these results with the numerical simulation of n-WE in order to assess the error made in approximating n-WE results with those of CWE. n-WE simulations are performed by straightforwardly generating the reduced density matrix using Wishart matrices (\ref{eq:defW}), calculating $Y$ and $\xi$ according to Eq.(\ref{eq:nCWE}), using theoretical values for GUE ensemble of $g_t=J_1(2t)/t$, where $J_1$ is the Bessel function, and $h_t=g_t^2$, both holding for large $NM$ (for large $NM$ used n-CWE and n-WE results are indistinguishable on the scale of figures).  
 
Three different cases of initial states will be considered: (i) product initial state, (ii) entangled initial state with only two nonzero Schmidt coefficients and (iii) entangled initial state with all $N$ nonzero Schmidt coefficients. 

\subsection{Product initial state}
\begin{figure}[t!]
  \includegraphics[angle=0,width=0.9\textwidth]{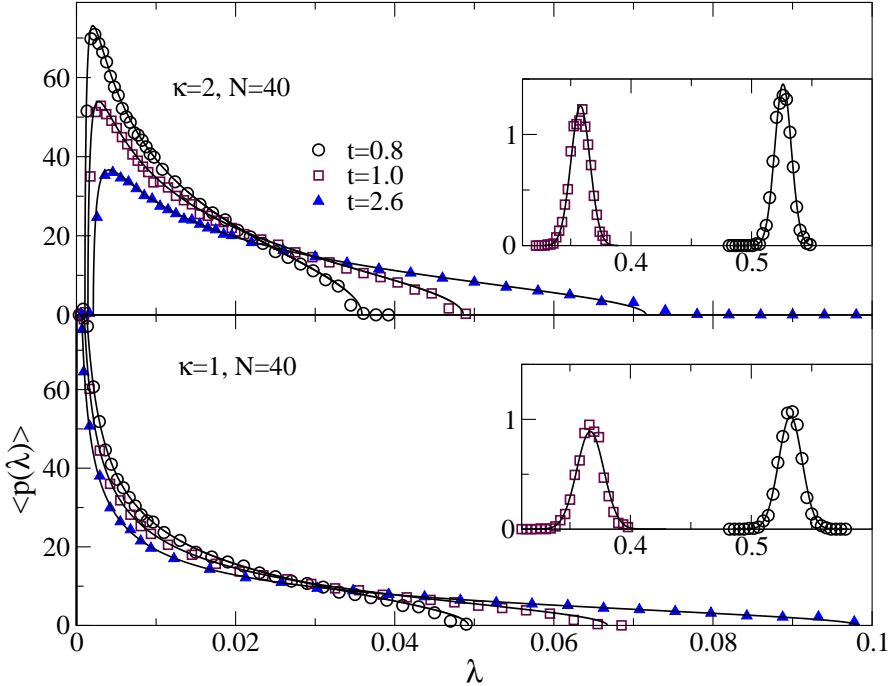}
  \caption{(Color online) Density of eigenvalues for RDMM (points) and theory (lines), Eq.(\ref{eq:lam1},\ref{mp}), for different values of $t$ and a product initial state. In the insets we show the largest eigenvalue, whose density is Gaussian with theoretical average (\ref{eq:lam1}) and variance (\ref{eq:var1}).}
  \label{fig:den}
\end{figure}
We start with the case (i), product initial state. In Fig.~\ref{fig:den} we show the eigenvalue density of the RDMM obtained by numerical simulation. These are compared with the theory for the bulk density (\ref{mp}) in the main plot and Gaussian distribution with the mean given by (\ref{eq:lam1}) and fluctuations by (\ref{eq:var1}) in the insets. Good agreement can be seen between the theory and RDMM results. We can see that the leading order of the CWE theory, giving a rescaled Mar\v cenko-Pastur density in the bulk, agrees well with the RDMM simulations, apart from small discrepancy around the edges of the spectrum. Theory for n-WE (data not shown) would be on the scale of the plot indistinguishable from the CWE theory, except near the bulk edges, where it would correctly account for small deviations of RDMM from the Mar\v cenko-Pastur density.

\begin{figure}[ht!]
  \includegraphics[angle=-90,width=0.8\textwidth]{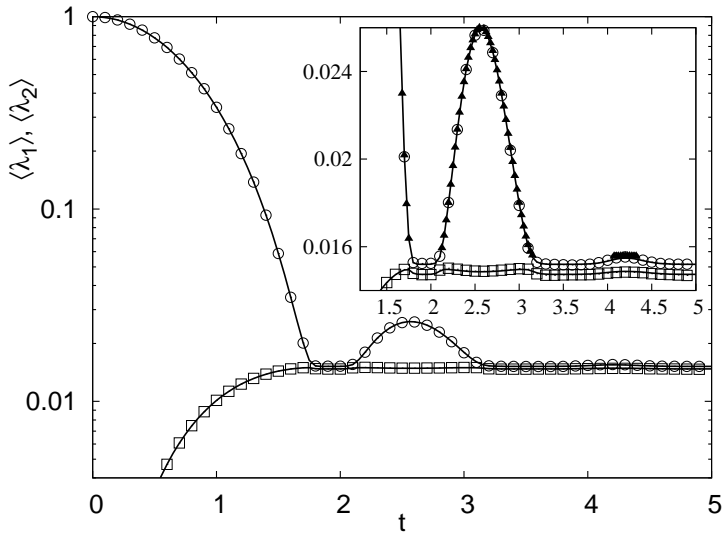}
  \caption{Average values of two largest eigenvalues for the RDMM simulation (points) and n-WE simulation (full curves) for an initial product state. The inset shows the ``collision'' between $\lambda_1$ and the bulk (filled triangles are theory (\ref{eq:lam1}) obtained for CWE). All is for $N=256, \kappa=1$.}
  \label{fig:lam01k1}
\end{figure}

\begin{figure}[ht!]
  \includegraphics[angle=-90,width=0.8\textwidth]{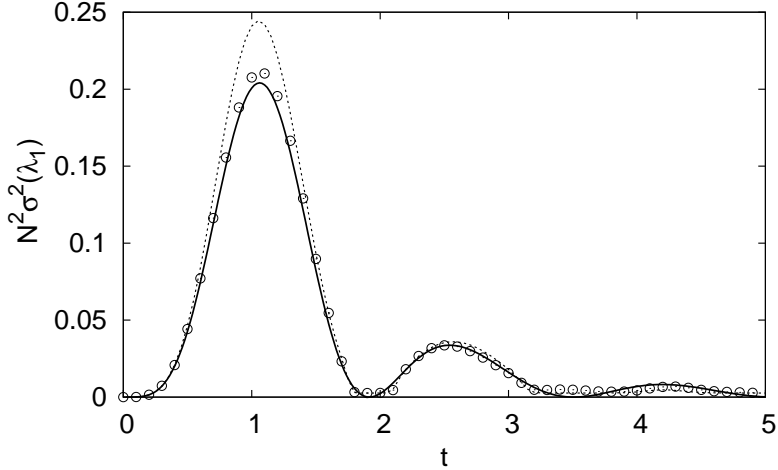}
  \caption{The variance of $\lambda_1$ for RDMM simulation (circles), theory Eq.~(\ref{eq:var1}) (full curve) and a fixed-trace n-WE simulation (dotted curve). All is for $N=256, \kappa=1$.}
  \label{fig:var1}
\end{figure}

Next, we proceed by a more detailed comparison of $\ave{\lambda_1}$. For an uncorrelated WE and large matrices $\ave{\lambda_j}$ have been calculated in Ref.~\cite{znidaric:07b}. In Fig.~\ref{fig:lam01k1} we show results of numerical simulation of RDMM, numerical simulation of n-WE, and analytical expression (\ref{eq:lam1}) obtained from CWE.
We can see a nice agreement between all three. In Fig.~\ref{fig:var1} similar comparison is made for the variance of the largest eigenvalue.

\begin{figure}[ht!]
  \includegraphics[angle=-90,width=0.8\textwidth]{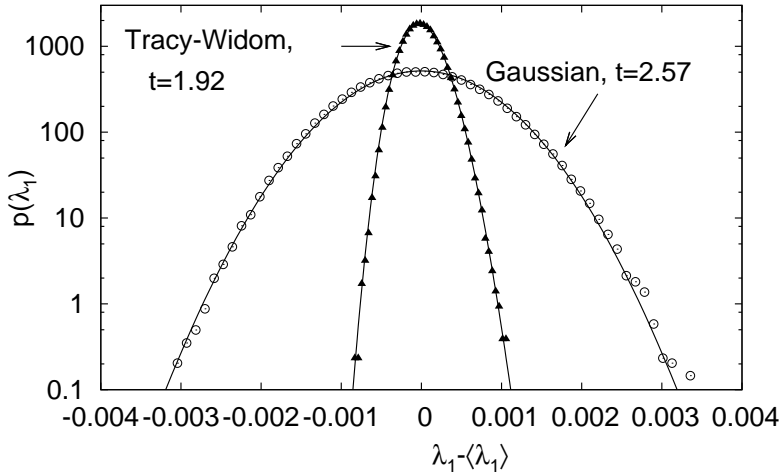}
  \caption{The distribution of the largest eigenvalue for n-WE (points) and theoretical Gaussian/Tracy-Widom distributions (full curves), indicating a phase transition with time. All is for $N=256, \kappa=1$.}
  \label{fig:phase}
\end{figure}

In the plot of $\ave{\lambda_1}$ and $\ave{\lambda_2}$, Fig.~\ref{fig:lam01k1}, one can nicely see an interesting ``collision'' between the largest isolated eigenvalue and the bulk, shown by the 2nd largest eigenvalue at its edge. Collision does not result in an exact degeneracy (see for instance the inset in Fig.~\ref{fig:lam01ic1}) but only in a very close encounter, with $\ave{\lambda_1}-\ave{\lambda_2} \sim 1/N^2$. Collisions happen approximately at times when $J_1(2t)$ has zeroes. Few things can be observed. First, there are interesting ``ripple'' effects at the time of collision (see Figs.~\ref{fig:lam01k1} and \ref{fig:lam01ic1}). More interestingly, at the time of the collision there is a phase transition in the distribution of the largest eigenvalue~\cite{Baik:05,Peche:06} from a Gaussian before the collision, to a Tracy-Widom distribution~\cite{TW} at the collision. This is demonstrated in Fig.~\ref{fig:phase}. Opposed to phase transitions in the distribution itself
 ~\cite{nadal:10} we here have a cascade of phase transitions in time. Note that the frequency of these collisions is proportional to the spectral span ($4$ for our GUE normalization), which is the fastest time scale in the system. 

\subsection{Entangled initial state}
\begin{figure}
  \includegraphics[angle=-90,width=0.8\textwidth]{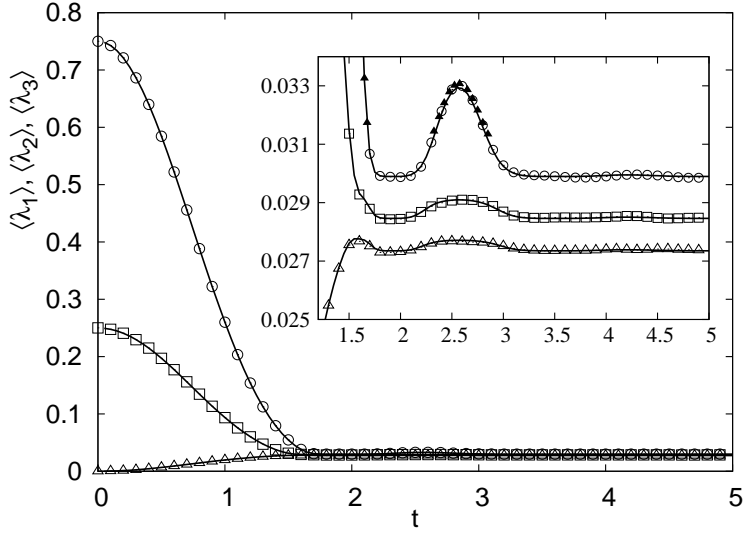}
  \caption{Three largest eigenvalues for $N=M=128$ and initial entangled state with two nonzero eigenvalues $\lambda_1=3/4$ and $\lambda_2=1/4$. Points are RDMM simulation, full curves n-WE theory obtained by numerical simulation and filled triangles in the inset Eq.(\ref{eq:lam1}) with $r=\frac{3}{4}h_t^2$.}
  \label{fig:lam01ic1}
\end{figure}

Next, we proceed to show results for two entangled initial states. The first one we choose is an entangled state with only two nonzero Schmidt coefficients, $\ket{\psi(0)}=\sqrt{1/4}\ket{0}\ket{0}+\sqrt{3/4}\ket{1}\ket{1}$. The eigenvalue density again agrees perfectly with n-WE simulations (data not shown). We only show a plot of three largest eigenvalues, Fig.~\ref{fig:lam01ic1}. One can again see that n-WE describes our RDMM well and that the largest eigenvalue is given by the analytical expression (\ref{eq:lam1}). 

\begin{figure}[ht!]
  \includegraphics[angle=0,width=0.8\textwidth]{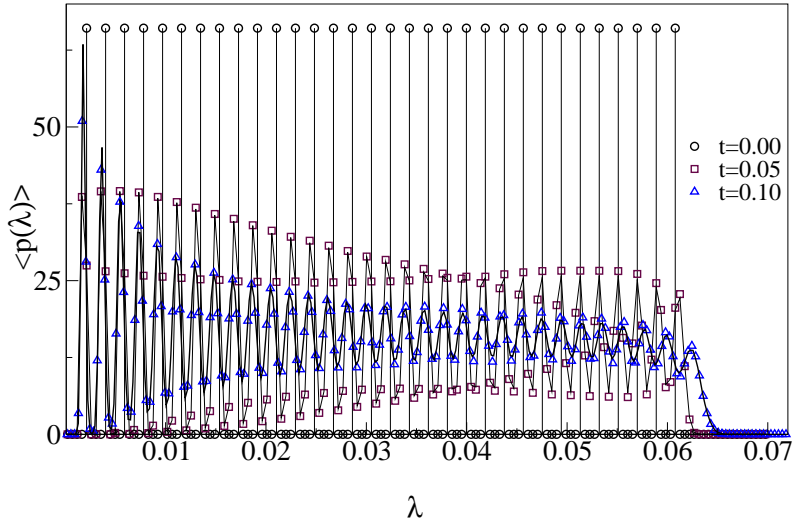}
  \caption{(Color online) Density of eigenlevels for $N=32$, $\kappa=1$ and initial state with $\lambda_{j}=2j/N(N+1)$ for $j=1,...,\,N$; points are RDMM simulation, lines are n-WE theory. Short time behavior showing transition from a ``picket-fence'' spectrum is shown. In Fig.~\ref{fig:spike2} long time behavior is shown.}
  \label{fig:spike1}
\end{figure}

\begin{figure}
  \includegraphics[angle=0,width=0.8\textwidth]{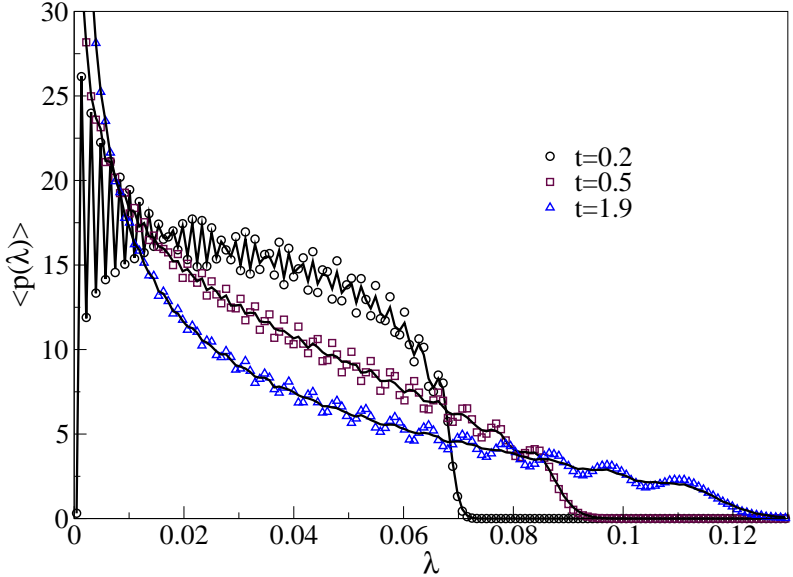}
  \caption{(Color online) Long time behavior of the same data as shown in Fig.~\ref{fig:spike1}. At $t\approx 1.9$, the first zero of $g_t \propto J_1(2t)/t$, correlations in the Wishart ensemble (\ref{eq:nCWE}) are zero and the density is described by the Mar\v cenko-Pastur form.}
  \label{fig:spike2}
\end{figure}

Finally, we choose a very entangled initial state, in which all Schmidt coefficients are nonzero, $\ket{\psi(0)}=\sum_j \sqrt{2j/N(N+1)} \ket{j}\ket{j}$. The eigenvalue density is shown in Figs.~\ref{fig:spike1}, ~\ref{fig:spike2} respectively for short time and for long time, again displaying an agreement with the n-WE theory. The oscillatory behaviour in Fig. ~\ref{fig:spike2} is due to the small system size. 

\section{Convergence to random states}
\label{sec:random}
Because purity can be used as an indicator of convergence to random states our exact results for purity in RDMM enable us to comment on the convergence rate towards random states. The Eq.(\ref{eq:I}) tells us that $I(t)-I_{\rm r} \approx g_t^4$. The convergence to a random state is reached with $1/N$ accuracy (which is equal to the scaling of $I_{\rm r}$ with $N$), $I(t)-I_{\rm r} \sim 1/N$, when we have $g_t \sim 1/N^{1/4}$. Because $g_t$ is the Fourier transformation of the spectral density its decay time is given by the inverse of the width of the spectral density. If the width is independent of the system size, as is the case for GUE, the decay time scales as $\sim 1$. If $g_t$ would be an exponential (Lorentzian eigenlevel density), such small value of $g_t$ would be reached after time $\tau \sim \log{N}\equiv n$. In a dynamical system one can decompose the propagator $U^t=\exp{(-\ii H t)}$ using e.g. Suzuki-Trotter formula into basic two-qubit operations. In a nearest-neighbor coupled chaotic system the number of two-qubit operations in a decomposition of a short-time propagator $U^{\Delta t}$ scales with the number of qubits $n$ linearly as $\sim n$. Decomposition up to time $\tau \sim n$ demands of order $\sim n^2$ two-qubit operations. Our result therefore indicates the same scaling with the number of two-qubit gates as in random circuits~\cite{circLin}. 

However, if $g_t$ is not an exponential there are two interesting issues: (i) if $g_t$ has zeroes (as is for instance the case for GUE) one can reach random states already after a much smaller time $\tau \sim 1$. Note that if the spectral span of eigenenergies is bounded one necessarily has zeroes in $g_t$. This is very promising as one could decrease convergence time by a factor of $n$ compared to known protocols~\cite{circLin}. (ii) On the other hand, for a bounded spectrum one will have a singular behavior at the spectral edge, for instance a square-root singularity for GUE, resulting in a very slow algebraic decay of the envelope of the oscillating $g_t$. This will result in an exponentially large time $\sim 2^n$ after which these oscillations die out. A bounded spectrum therefore has advantages and disadvantages.

It is an interesting open question how these issues are reflected in a concrete dynamical system. What is clear from our discussion is that what matters is the Fourier transformation of the spectral density, as well as the spectral span, which trivially rescales all times given in our work. ``Engineering'' an appropriate spectral density one could speed-up the convergence.

\section{Conclusion} 
\label{sec:conc}
Using numerical simulations and asymptotic arguments we showed that one can describe the statistics of the reduced density matrix during unitary evolution by a complex Hamiltonian with a noncentral Wishart ensemble. We have analytically calculated purity, the average largest eigenvalue and its variance. Interesting collisions between separated eigenvalues and a random-matrix-like bulk occur as time progresses, leading to phase transitions in the distribution of these eigenvalues. We discuss convergence time for a large class of protocols for generation of random states, also suggesting a new way to speed-up these protocols. The results presented open many new questions in physics as well as in mathematics. For instance, it would be interesting to study generalization to orthogonal or symplectic symmetry. 

\ack

We acknowledge discussions with T.~H.~Seligman and support by the project 79613 by CONACyT, Mexico, and IN114310 by UNAM.
\vskip4mm
{\em NOTE: after submitting this work two preprints have appeared~\cite{recent}, discussing convergence properties under random evolution.}

\appendix
\section*{Appendix}
\setcounter{section}{1}
\label{sec:appendix}
In this Appendix we sketch the steps needed in evaluating averages over unitary Haar measure, needed in calculation of average reduced density matrix (\ref{eq:avgrho}), average purity (\ref{eq:I}) and fluctuations of the largest eigenvalue (\ref{eq:var1}). They can also be found in previous works, for instance, in Refs.~\cite{Collins,znidaric:11}.

Let us write the propagator $U^t_{j\mu,k\nu}=V_{j\mu,m\lambda} {\rm diag}\{\exp{(-\ii E_{m\lambda} t)}\} V^*_{k\nu,m\lambda}$ in terms of eigenvectors $V_{j\mu,k\nu}$ and eigenenergies $E_{m\lambda}$. Summations are implied over repeated indices. We use notation with double indices, for instance $j\mu$, where Latin indices are for the central subsystem of interest, being of dimension $N$, while Greek indices enumerate the environmental basis states, being $M$ in number, over which we trace in order to obtain the reduced density matrix. Expansion coefficients $A_{j,\nu}(t)$ of the state at time $t$ are therefore
\begin{equation}
A_{j,\nu}(t)=U_{j\nu,k\mu}^t A_{k,\mu}(0).
\end{equation}
To evaluate the average expansion coefficient we therefore need
\begin{equation}
\ave{A_{j,\nu}(t)}_{\rm U}= \ave{V_{j\nu,m\lambda}V^*_{k\mu,m\lambda}}_{\rm U}\, {\rm e}^{-\ii E_{m\lambda}t} A_{k,\mu}(0).
\end{equation}
Using the well known result for the Haar average of a product of two matrix elements,
\begin{equation}
\ave{V_{j\nu,m\lambda}V^*_{k\mu,m\lambda}}_{\rm U}=\frac{1}{{\cal N}} \delta_{j\nu}^{k\mu},
\end{equation}
we immediately get the Eq.(\ref{eq:ft}). We abbreviated ${\cal N}\equiv NM$ and use a short notation for a product of Kronecker delta symbols, $\delta_{j_1 j_2\ldots}^{\nu_1 \nu_2 \ldots}=\delta_{j_1,\nu_1} \delta_{j_2,\nu_2}\cdots$.

To calculate the average value of the reduced density matrix, obtained as $\rho(t)=A(t)A^\dagger(t)$, we need the average of a product of four matrix elements of $V$ (two from each $A$). They are equal to
\begin{equation}
\fl
\ave{V_{a_1b_1}V_{a_2b_2}V^*_{a_3b_3}V^*_{a_4b_4}}_{\rm U}=\frac{1}{{\cal N}^2-1}\left(\delta_{a_1b_1a_2b_2}^{a_3b_3a_4b_4} +\delta_{a_1b_1a_2b_2}^{a_4b_4a_3b_3} -\frac{1}{{\cal N}}\left[ \delta_{a_1b_1a_2b_2}^{a_3b_4a_4b_3} +\delta_{a_1b_1a_2b_2}^{a_4b_3a_3b_4}\right] \right) ,
\end{equation}
where we used a short notation $V_{a_ib_i}\equiv V_{j_i\nu_i,k_i\mu_i}$ with a double-index $a_i\equiv j_i\nu_i$ and $b_i\equiv k_i\mu_i$. Using this formula in the expression for $\rho(t)$, summing over all repeated indices, and finally simplifying the resulting products of Kronecker deltas, one arrives at the average reduced density matrix as given in Eq.~(\ref{eq:avgrho}).

For purity, which is quadratic in the reduced density matrix, we have to average over product of 8 $V$s. For this one needs the Haar average over a product of 8 matrix elements. Exact formula is rather lengthy, however, using the result in Ref.~\cite{Collins}, averages over unitary group can be written in the following compact form
\begin{equation}
\fl
\ave{V_{a_1b_1}\cdots V_{a_pb_p} V^*_{a'_1b'_1} \cdots V^*_{a'_pb'_p}}_{\rm U}=\sum_{\sigma,\tau \in S_p} \delta_{a_1\ldots a_p\, b_1\ldots b_p}^{a'_{\sigma_1} \ldots a'_{\sigma_p}b'_{\tau_1}\ldots b'_{\tau_p}}\,{\rm Wg}({\cal N},\sigma \tau^{-1}),
\label{eq:W}
\end{equation}
where the summation is over all permutations $\sigma$ and $\tau$ of $p$ elements. Weingarten function ${\rm Wg}$ is a rational function in size ${\cal N}$ and depends only of the cycle shape of the permutation $\sigma \tau^{-1}$. The whole evaluation of unitary Haar averages therefore boils down to knowing values of the Weingarten function. For $p=2$ and $p=3$ they are given in Ref.~\cite{Collins}, for $p=4$, needed in purity calculation, they have been calculated in Ref.~\cite{znidaric:11}. For permutations of $p=4$ elements there are 5 different values of Weingarten function (i.e., 5 different cycle shapes), while the sum in Eq.(\ref{eq:W}) results in a sum of $(4!)^2=576$ products of Kronecker deltas. A sheer number of terms and the fact that one has to sum all these terms over repeated indices, taking into account also exponential factors involving energies, makes calculation rather long and tedious. After summation and simplification many terms actually result in the same expression so that the final results can be compactly written as given in Eq.(\ref{eq:I}).


\begin{thebibliography}{}
\bibitem{footnote2} At $t=0$ the fluctuations for CWE do not go to zero, as they do in the density matrix. Fixed-trace n-WE, however, has almost the correct variance, see Fig.\ref{fig:var1}.
\end{thebibliography}
\end{document}